# New results on the Coma cluster: revealing the primary component

F. Durret[1,2], A. Biviano[3], D. Gerbal[1,2], O. Le Fèvre[2], C. Lobo[1,4], A. Mazure[5], E. Slezak[6]

[1] *Institut d'Astrophysique de Paris, CNRS, 98bis Bd Arago, 75014 Paris, France*
[2] *DAEC, Observatoire de Paris-Meudon, 92195 Meudon Cedex, France*
[3] *Leiden Sterrewacht, Postbus 9513, Niels Bohrweg 2, 2300 RA Leiden, The Netherlands*
[4] *Centro de Astrofísica, Rua Campo Alegre 823, 4150 Porto, Portugal*
[5] *LAS, Les Trois Lucs, B.P. 8, 13376 Marseille Cedex, France*
[6] *Observatoire de la Côte d'Azur, B.P. 229, 06304 Nice Cedex 4, France*

Recent observations of the Coma cluster of galaxies in its central region have provided approximately 250 new redshifts - allowing a good membership criterion to be established for brighter galaxies - and magnitudes for 8000 objects in the same region derived from photometric data complete up to $V_{26.5}=22.5$. A thorough structural study of the galaxy distribution of the cluster galaxies, jointly with an X-ray wavelet analysis, and with a kinematical analysis of the velocity distribution allows us to uncover a primary body of the cluster, with evidence for a velocity gradient.

1. INTRODUCTION

The Coma cluster had been considered for a long time as the archetype of relaxed clusters, until the discovery in the 80's of substructures in many clusters, including Coma. In its very center lie two subclusters, centered on the well-known giant galaxies NGC4874 (a rich group) and NGC4889 (less rich). Another rich group around NGC4839 in the south-west is observed at optical as well as X-ray wavelengths. Coma is now the archetype of rich clusters, endowed with subclusters, and thus not really relaxed.

The purpose of this paper is to show that the situation is more intricate: the formation of clusters is a process still in action; therefore, we have an already formed (primary) cluster that is continuously aggregating matter by the infall of galaxies, groups, and even clusters, filled or not with gas, and with or without unseen matter.

We have tried to detect such a primary component in Coma.

## 2. OPTICAL DATA

### 2.1. Definition of the sample.

Galactic positions and $b_{26.5}$ magnitudes (hereafter simply noted as b) are taken from the GMP catalog (Godwin et al. 1983). Redshift information is the one described in Biviano et al. 1995b.

In order to perform our analysis we limit this set of data to all the galaxies located in the circle centered in the GMP center with 1500 arcsec radius.There are 544 galaxies in this sample up to magnitude b = 20.0. Velocities are available for 316 of these galaxies.

The set of 383 *cluster members* in this region which we used to produce the luminosity function is defined by: i) the redshift information: 255 galaxies belong to the cluster as determined by their redshift: their velocities are comprised in the interval [2990,10000] km s$^{-1}$. We are thus 67% complete in redshift up to b = 20.0; ii) the colour - magnitude band (see Biviano et al. 1995a for a more detailed description).

### 2.2. Structural analysis

We used the adaptive kernel method (e.g. Pisani 1993, Merritt & Tremblay 1994) in 2-D to make isopleth charts of the optical data. Figs. 1, 2 and 3 show the results when this technique is applied, respectively, to all *cluster members*, the *bright* (b≤ 17.0, 122 objects) and *faint* (b> 17.0, 261 objects) samples.

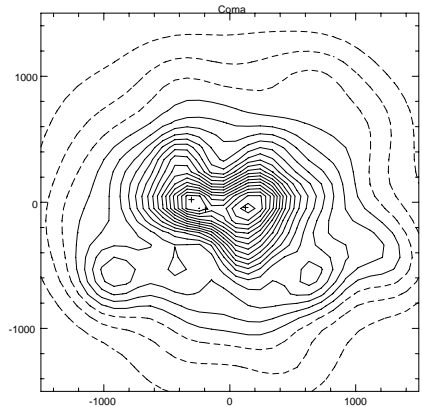

Fig. 1. Density kernel map, all galaxies.

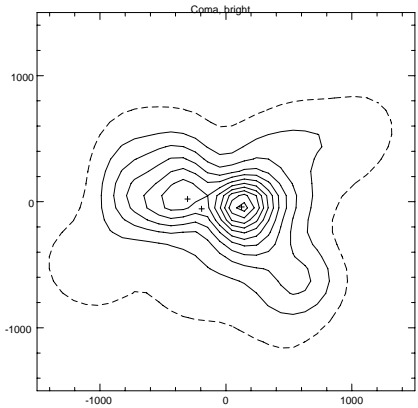

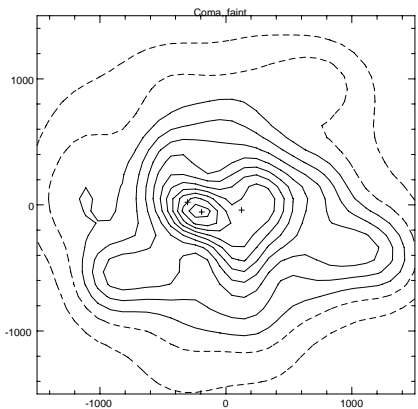

Figs. 2 and 3. Density kernel maps, bright galaxies (top) and faint galaxies (bottom).

- The well known picture of the Coma center is well reproduced when all galaxies are considered: we can easily see the contours revealing the presence of subclusters around the giant galaxies NGC 4889 and NGC 4874 located east and west of the GMP center, as indicated in the figure by the crosses. Further away from the center contours become smoother.
- For the *bright sample* chart, luminosity segregation stands out, especially around NGC4874 (right cross indicates its position), enhancing substructure patterns.
- As for *faint* galaxies, they trace a more regular cluster structure where both previous central clumps are no longer relevant features. The peak in this distribution is located between the position of the two "monster" galaxies (the middle cross, in the figure) and not far from the GMP center.

  The outer parts of the cluster core remain quite similar in all three charts.

## 3. X-RAY DATA AND ANALYSIS

We have used images (kindly provided by S. White) taken with the ROSAT PSPC and described in White et al. (1993). We have applied a wavelet analysis to these images (see Slezak et al. 1994); this method allows us to remove noise and to detect features at the level of significance we want. We display in figs. 4 and 5 the analyses at two successive scales, where one may clearly see: 1) the emission coming from the groups around the two giant galaxies; two other features *with no optical couterpart* are visible; 2) an intermediate peak between the two peaks centered on NGC 4874 and NGC 4889 (previously pointed out by White et al. 1993); 3) a filamentary structure along the brink of NGC 4874. On an even larger scale we find again the general behaviour of optical data put in evidence in fig. 1, indicating that

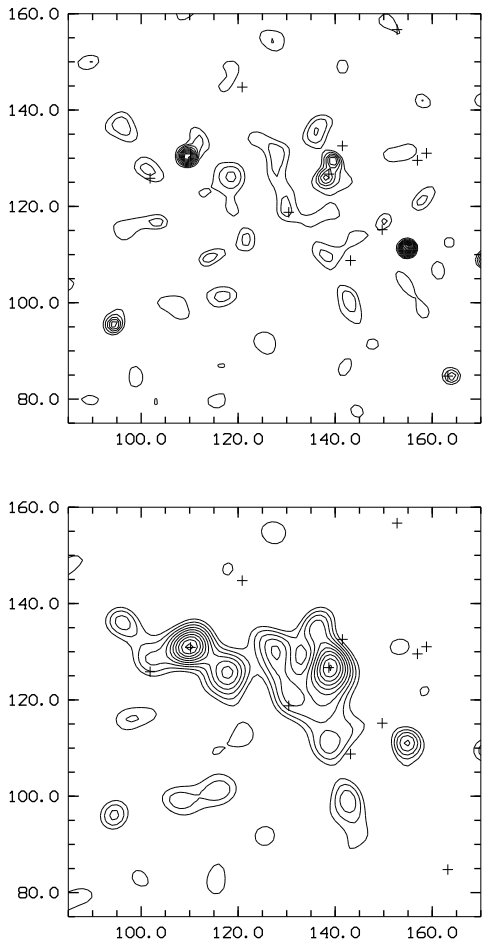

Figs. 4 and 5. Isopleths for the intermediate (top) and large scale (bottom) wavelet planes; crosses indicate the ten brightest galaxies in X-rays, according to Dow & White (1995).

## 4. KINEMATICAL PROPERTIES
### 4.1. Velocity histograms

We studied the velocity distributions of the three samples described in the last section. Table 1 illustrates the results: columns 2 and 3 show the mean velocity and the velocity dispersion $\sigma$ obtained by a biweight analysis. The velocity distribution of galaxies in the central region of Coma is not very different from a gaussian.

TABLE 1. Characteristics of the velocity samples.

| sample | mean | $\sigma$ | number |
|---|---|---|---|
| all | 6901 | 1142 | 255 |
| bright | 6827 | 1121 | 121 |
| faint | 6967 | 1161 | 134 |

### 4.2. A velocity gradient.

We investigated the existence of a velocity gradient. We found no gradient either for the general sample as previous authors had already concluded (Gregory 1975) or for the *bright sample*, but did find a significant one for the *faint sample* alone.

The method we used was the following:
- We looked for a significative *direction* and calculated the correlation between the velocity and the abscissa of galaxy positions. We found a maximum correlation for the direction given by $\theta_0 = 75$ degrees.
- We have constructed the profile of the running mean velocity in the direction $\theta_0$ for the *faint sample* (fig. 6). The significance of this gradient is 0.972 according to a non-parametric Kendall correlation. The result indicates large variations of the $<V>$ amplitude : from +300 to -300 throughout the x-axis on the plot, typical of a rigid body rotation, or possibly a shear, in the central part. As we only possess velocity information along the line of sight, we cannot distinguish between these two physical behaviours.

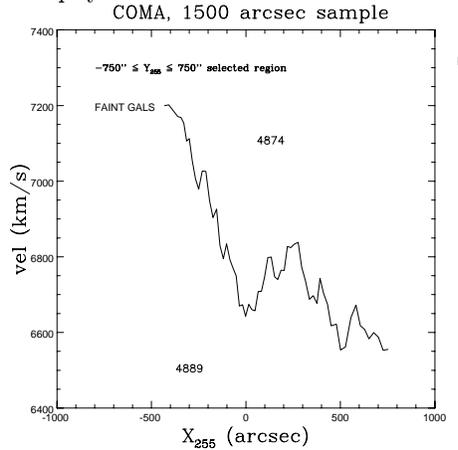

Fig. 6. Running mean velocity (each point is the average of 50 velocities) in the *faint sample*, along the 255 degrees rotated X axis, by selecting only galaxies within a zone +/- 750" on the rotated y-axis.

Another way to see this gradient is by displaying the mean velocity kernel map (fig. 7). We can easily see the gradient mentioned above (although we also have some strange features in the upper part of the figure). With this new result in mind, we re-checked the velocity histogram for the *faint sample* after subtraction of the "rotation" velocity. The result is a slightly less asymetrical distribution but with a larger kurtosis value (distribution more highly peaked) than before.

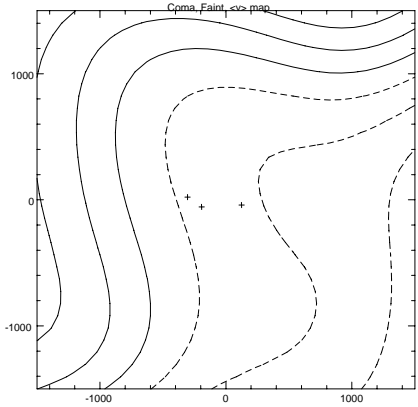

Fig. 7. Mean velocity kernel for *faint* galaxies without both central groups; contour levels are spaced by 100 km s$^{-1}$.

4.3. Density maps for different velocities

In figures 8 and 9 we show charts of iso-densities, again with the adaptive kernel method, here applied to two different subsets of our 1500 arcsec sample of galaxies: one subset contains all galaxies with velocities v > 6600 km s$^{-1}$ while the other subset contains those with v < 6600 km s$^{-1}$.

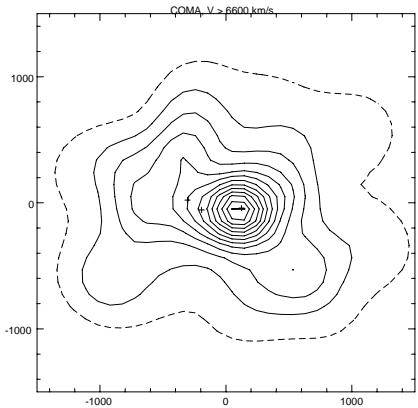

Fig. 8. Density kernel map (kernel size = 1 silverman) for all galaxies with v > 6600 km s$^{-1}$.

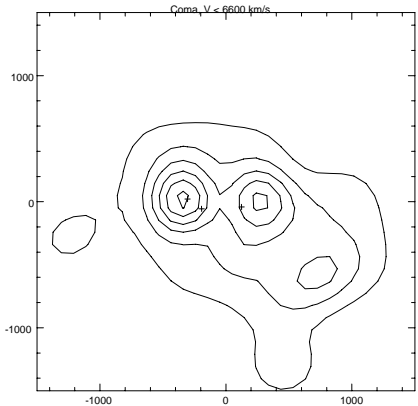

Fig. 9. Density kernel map (kernel size = 1 silverman) for all galaxies with v < 6600 km s$^{-1}$.

The situation appears complex but one can see a general trend for galaxies with v > 6600 km s$^{-1}$ to cluster mostly around NGC 4874, and for the others to cluster mostly around NGC 4889.

## 5. DISCUSSION AND CONCLUSIONS

The three kinds of behaviours described above (structural, X-ray features, kinematical) can be interpreted according to a general view of formation and evolution of clusters in general, and for the Coma cluster in particular. Coma is embedded in a filament that runs along the direction defined by Coma itself and its neighbour Abell 1367 (Fontanelli 1984, Slezak et al. 1988), but also lies at the border of the Great Wall, as can be seen in the ($\alpha$,z) plane where galaxy bubbles join their walls (Geller & Huchra 1989).

From the dynamical point of view, a main body would have formed by violent relaxation and mixing by the arrival and merging of groups. Successive accretion of galaxies or small groups following trajectories along these galaxy walls, as an ongoing process, have given place, in an ancient period of the history of this cluster, to this now relaxed primary component (West 1994). Notice that the definition of subgroups logically implies the actual existence of a background primary body from which substructures can stick out.

We provide evidence that this primary component exists in Coma, is well traced by the *faint* population of galaxies, and has kinematical properties consistent with a rotating rigid body. In fact, it is precisely with the *faint* galaxies that we have had a possibility of revealing the primary component because more luminous galaxies are affected by various phenomena which are not directly linked to the primary component (such as luminosity segregation, starburst events, etc...). Moreover, this relaxed component and the potential it generates is filled with X-ray emitting gas smoothly distributed as are galaxies. It is thus natural that the peak of this emission (intermediate maximum in the figure) coincides with the peak of the distribution of *faint* galaxies, emphasizing simultaneously the primary component bottom of the potential well.

The NGC 4874 and NGC 4889 groups have mean velocities in disagreement with the gradient. They seem to have arrived not long ago: dynamical friction of these groups travelling through the cluster may have not yet been efficient. The filament described above could be the signature of a compression wave due to the infall of the group on to the primary component. NGC 4874 is a more important group than NGC 4889 (Mellier et al. 1988, Escalera et al. 1992); it may possess an important X-ray gas component, and NGC 4839 is a well known subcluster now arriving at or perhaps having already having crossed Coma (Burns et al. 1994).

**Acknowledgements:** C. Lobo is fully supported by the BD/2772/93RM grant attributed by JNICT, Portugal.